\documentclass{article}
\PassOptionsToPackage{numbers, sort}{natbib}
\usepackage[preprint]{neurips_2026}

\usepackage[utf8]{inputenc}
\usepackage[T1]{fontenc}
\usepackage{hyperref}
\hypersetup{
    colorlinks=true,
    linkcolor=[RGB]{31,119,180},   
    citecolor=[RGB]{44,160,44},    
    urlcolor=[RGB]{214,39,40},     
}
\usepackage{url}
\usepackage{booktabs}
\usepackage{amsfonts}
\usepackage{makecell}
\usepackage{amsmath}
\usepackage{nicefrac}
\usepackage{microtype}
\usepackage{xcolor}
\usepackage{graphicx}
\usepackage{cleveref}
\usepackage{wrapfig}   
\usepackage{adjustbox}
\usepackage{algorithm}
\usepackage{algpseudocode}
\newcommand{\sgname}{\textsc{SkillGuard}}
\newcommand{\dbname}{\textsc{DriftBench}}
\newcommand{\skill}{\mathcal{S}}
\newcommand{\contract}{\mathcal{C}}
\newcommand{\drift}{\mathcal{D}}

\title{Skill Drift Is Contract Violation: \\ Proactive Maintenance for LLM Agent Skill Libraries}

\author{%
  Linfeng Fan$^{1}$ \quad Yuan Tian$^{1}$ \quad Ziwei Li$^{2}$ \quad Zhiwu Lu$^{1,*}$ \\[4pt]
  $^{1}$Gaoling School of Artificial Intelligence, Renmin University of China, Beijing, China \\
  $^{2}$King Abdullah University of Science and Technology, Thuwal, Saudi Arabia \\[4pt]
  \texttt{\{%
    \href{mailto:2023200424@ruc.edu.cn}{2023200424},
    \href{mailto:tianyuan2004@ruc.edu.cn}{tianyuan2004},
    \href{mailto:luzhiwu@ruc.edu.cn}{luzhiwu}%
  \}@ruc.edu.cn,\;
  \href{mailto:ziwei.li@kaust.edu.sa}{ziwei.li@kaust.edu.sa}} \\[2pt]
  {\small $^{*}$Corresponding author.}
}
\begin{document}

\maketitle


\begin{abstract}
LLM agents increasingly rely on reusable skill libraries, but these skills silently decay as the external services, packages, APIs, and configurations they reference evolve. Existing monitors detect such changes at the wrong granularity: they observe values, not the role those values play in a skill. A version string in a comment is noise; the same string in a pinned dependency is an operational obligation. We formulate skill drift as contract violation and introduce \sgname{}, which extracts executable environment contracts from skill documents and validates only those role-bearing assumptions against known or live conditions. This distinction turns noisy monitoring into a precision-first maintenance signal. Contract-free CI probes produce 40\% false positives, while \sgname{} raises zero false alarms over 599 no-drift and hard-negative cases (Wilson 95\% CI $[0,0.6]\%$). In known-drift verification, \sgname{} achieves 100\% precision and 76\% recall with the strongest backbone; in a pre-registered study over 49 real skills, it discovers live drift with 86\% conservative precision. Violated contracts also make repair actionable, improving one-round success from 10\% without localization to 78\%. We release \dbname{}, an 880-pair benchmark for skill degradation.
\end{abstract}


\section{Introduction}
\label{sec:intro}

LLM agents are moving from one-off task solving~\cite{besta2024graph,wang2023plan,park2023generative,yao2022react,raad2024scaling} to long-lived skill libraries~\cite{xu2026agent,xia2026grasp,wang2023voyager,tagkopoulos2025skillflow,zhou2026memento,liu2026skillforge}: reusable procedures for deploying services, writing tests, configuring infrastructure, and interacting with external tools. This shift makes maintenance a first-order reliability problem. A skill may be correct when written, yet fail later because the environment it assumes has changed. APIs migrate, packages deprecate interfaces, URLs move, configuration defaults change, and authentication mechanisms evolve~\cite{fruntke2025automatically}. The question is therefore not only how agents acquire useful skills, but how those skills remain valid after deployment.

\begin{figure}[ht]
    \centering
    \includegraphics[width=0.88\linewidth]{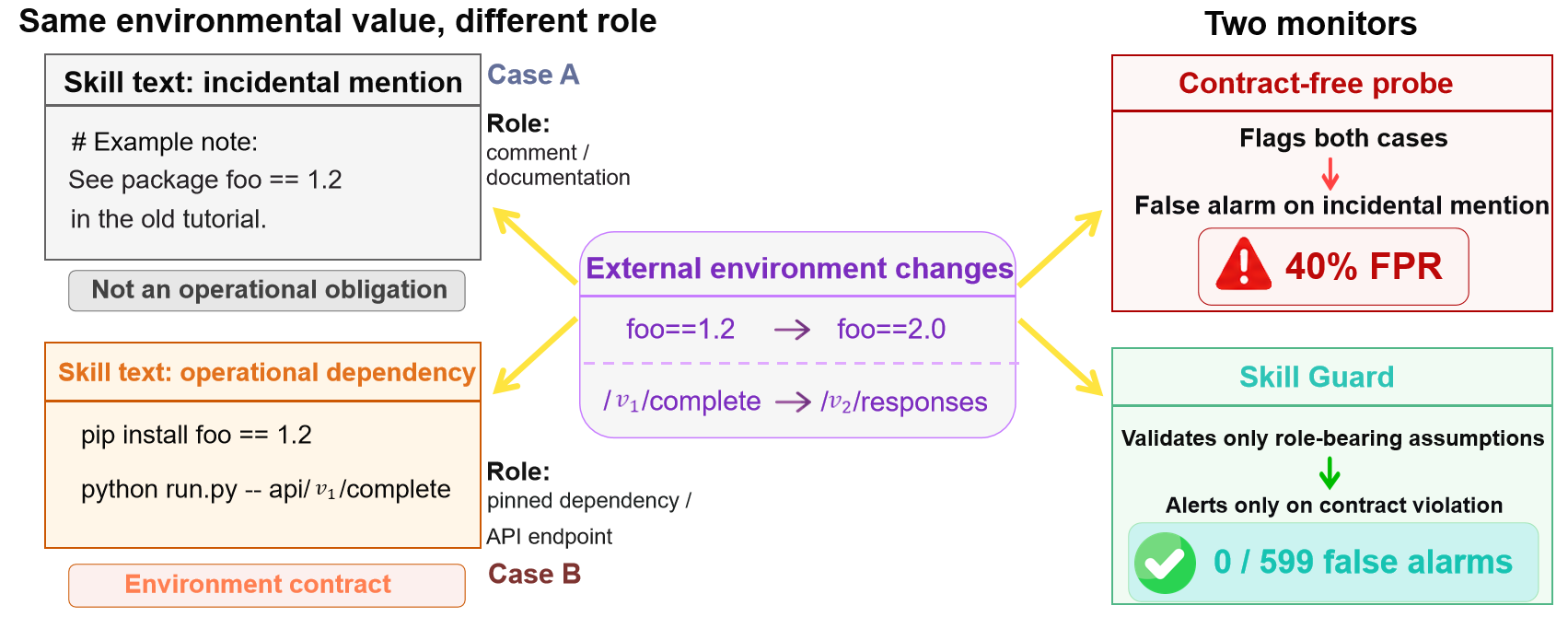}
    \caption{
    \textbf{Skill drift is role-dependent.}
    Raw environmental monitoring treats every changed URL, version, or configuration value as potentially relevant.
    \sgname{} instead distinguishes incidental mentions from operational obligations.
    This granularity explains the main empirical gap: contract-free CI probes produce 40\% FPR, while \sgname{} raises zero false alarms over 599 no-drift and hard-negative cases.
}
    \label{fig:contract_violation}
\end{figure}

Existing monitors are poorly matched to this problem because they observe environmental values rather than the role those values play in a skill. Package registries can report version bumps, URL probes can find redirects, and CI checks can expose broken endpoints. These signals say that the world changed; they do not say whether the change matters for a particular skill. A version string in a comment is noise; the same string in a pinned dependency is an obligation. A documentation redirect may be harmless; an API endpoint redirect may break the procedure. As \Cref{fig:contract_violation} illustrates, monitoring raw values forces a bad trade-off: alert on incidental change and create false alarms, or filter aggressively and miss drift embedded in natural-language procedures.

The key observation behind this paper is that skill drift is role-dependent. Drift is not a property of an external value alone, but of the role that value plays in the skill. We formalize this view by treating skill drift as contract violation. A skill implicitly carries environment contracts: testable assumptions about services, packages, APIs, configurations, schemas, and credentials that must remain valid for the skill to execute as intended. The monitoring target is therefore not ``what changed in the world?'' but ``which role-bearing assumptions of this skill no longer hold?''

We introduce \sgname{}, a precision-first monitor that operationalizes this contract view. Given a skill document, \sgname{} extracts executable environment contracts and validates only those contract-bearing assumptions against known or live conditions. This separates operational dependencies from incidental mentions before probing the environment. Failed contracts also provide localized repair evidence: instead of asking an LLM to rewrite a stale skill from scratch, \sgname{} identifies the specific assumptions that must be updated.

We evaluate this view through three claims. First, contracts suppress false alarms: CI-style probes without contracts produce 40\% false positives, whereas \sgname{} raises zero false alarms across 599 no-drift and hard-negative cases (Wilson 95\% CI $[0\%,0.6\%]$). Second, precision does not come from ignoring drift: in known-drift verification, the strongest backbone achieves 100\% precision and 76\% recall. Third, the framing transfers beyond closed-world specifications: in a pre-registered study over 49 real skills, \sgname{} discovers live drift with 86\% conservative precision. Violated contracts also make repair local rather than open-ended, improving one-round success from 10\% without localization to 78\%.

Our contributions are:
\begin{enumerate}
    \item \textbf{A contract-violation framing for skill drift.} We identify the granularity mismatch in skill maintenance: useful monitoring must distinguish operational environment assumptions from incidental environmental mentions.
    \item \textbf{\sgname{}, a contract-based maintenance monitor.} \sgname{} extracts executable environment contracts from skill documents, validates them against known or live conditions, and uses violations to localize repair.
    \item \textbf{\dbname{} and empirical evidence.} We introduce an 880-pair benchmark with 174 generated drifts, 107 real-world drifts, and 599 negative controls, and show that contract granularity yields precision-first detection, live discovery, and targeted repair without claiming full skill verification.
\end{enumerate}
\section{Related Work}
\label{sec:related}

\paragraph{Agent skills, skill benchmarks, and skill auditing.}
LLM agents increasingly package procedural knowledge as reusable skills rather
than solving each task from scratch. Voyager~\cite{wang2023voyager} builds a
skill library through self-verification, and SkillFlow~\cite{tagkopoulos2025skillflow}
discovers reusable skills from agent trajectories. Recent work treats skills as
first-class artifacts: SkillsBench evaluates whether curated or self-generated
skills improve agent performance across heterogeneous tasks~\cite{li2026skillsbench},
and systematization work on agentic skills studies their representation,
composition, evaluation, update, and governance~\cite{jiang2026agenticskills}.
A closely related security line audits whether untrusted Agent Skills are safe
to load, using role-aware evidence extraction, semantic verification, and
adjudication to identify unsafe skill packages~\cite{lv2026skillguardrobust}.
These works establish the importance of the skill layer, but they target
different lifecycle questions: skill acquisition and usefulness, pre-load
security auditing, or governance. \sgname{} studies post-deployment maintenance:
whether a previously valid skill has become stale because its environment-facing
assumptions no longer hold.

\paragraph{Reactive repair and localization.}
Automated program repair localizes faults and patches code after failures are
observed~\cite{sepidband2026rgfl,bouzenia2025repairagent,chen2023teaching}.
Self-refine~\cite{madaan2023self} and Reflexion~\cite{shinn2023reflexion}
similarly improve model outputs through iterative feedback. These methods are
reactive: they require a failed execution, test, or trajectory before repair is
attempted. \sgname{} instead targets proactive maintenance. It detects violated
environment contracts before relying on an execution failure, and it uses the
failed contract as a localized repair signal. Accordingly, our repair result
should be read as evidence for localization, not as a claim of dominance over
stronger multi-round repair methods.

\paragraph{Contracts and specification mining.}
Executable contracts and specification mining extract assumptions from
programs, traces, or documentation~\cite{meyer2002applying,ernst2007daikon,bhardwaj2026agent,lahiri2026intent,chen2026codespecbench,doshi2026towards,dong2024building}.
Our use of contracts differs in both source and purpose. Agent skills often mix
operational dependencies with comments, examples, badges, prose, and partial
configuration snippets. The main difficulty is therefore not only extracting a
predicate to check, but deciding whether an environmental value should be
monitored at all. In \sgname{}, a contract is a role-bearing obligation:
incidental mentions are suppressed, while pinned dependencies, API endpoints,
schema fields, authentication mechanisms, and configuration values are
validated.

\paragraph{Dependency, API, and configuration-drift monitoring.}
Dependency update tools, API-evolution analyses, and DevOps monitors track
changes in structured software ecosystems~\cite{dig2006apis,fruntke2025automatically}.
Recent LLM-agent systems also study configuration drift in cloud and
infrastructure-as-code settings~\cite{abuzakuk2026riva}. These systems typically
assume structured manifests, telemetry, IaC specifications, or executable
environments. Agent skills are less structured: environment assumptions may be
embedded in natural-language instructions, shell snippets, examples, partial
configs, and documentation references. This changes the monitoring objective.
Probing every observed value creates false alarms, while probing only declared
manifests misses assumptions expressed in prose. \sgname{} addresses this gap by
validating role-bearing environment contracts rather than all observed values.

\paragraph{Agent and software-maintenance benchmarks.}
Agent benchmarks such as AgentBench~\cite{liu2023agentbench}, SWE-bench~\cite{jimenez2023swe},
and SWE-Skills-Bench~\cite{han2026swe} evaluate task-solving, software
engineering, and skill quality. These benchmarks are valuable, but they do not
isolate the maintenance failure studied here: a skill that was once valid can
degrade when the external environment changes. \dbname{} complements them by
pairing previously valid skills with generated drifts, LLM-free real-world
drifts, and hard negatives that change text or values without violating
operational contracts. This design tests both drift recall and false-alarm
resistance, the two sides of precision-first skill maintenance.
\section{\dbname{}: A Benchmark for Skill Degradation}
\label{sec:driftbench}

A benchmark for skill drift must test two failure modes simultaneously: missing stale operational assumptions and over-alerting on environmental changes that do not affect the skill. \dbname{} is built around this distinction. It includes controlled drift pairs for recall, real-world drifts for external validity, and no-drift and hard-negative pairs for false-alarm resistance. This design makes the benchmark aligned with the paper's central claim: skill drift depends on whether an environmental change violates a role-bearing contract, not on whether a value changed at all.

\paragraph{Drift taxonomy.}
We define eight drift types from software maintenance failures and real changelogs: \textbf{URL change}, \textbf{version bump}, \textbf{configuration change}, \textbf{API migration}, \textbf{deprecation}, \textbf{schema change}, \textbf{authentication change}, and \textbf{dependency update}. These categories cover both direct surface changes, such as moved endpoints or renamed imports, and less syntactically obvious changes, such as schema-field updates or authentication-mechanism changes.

\paragraph{Controlled drift construction and validation.}
\dbname{}  starts from 49 agent skills in SWE-Skills-Bench~\cite{han2026swe} across seven programming domains. Qwen3.6-Plus proposes 6--8 drift scenarios per skill, yielding 422 candidate pairs; validated candidates form a controlled evaluation split with 174 drift pairs. To reduce generator artifacts, we use cross-family validation: DeepSeek-R1 independently judges candidate pairs, disputed cases are checked against registry-backed evidence, and adjudication decisions are reviewed by the authors. The resulting controlled split reaches 99.6\% adjudicated validity.

\paragraph{LLM-free real-world split.}
Synthetic drift alone is not enough for a maintenance benchmark. We therefore collect 107 naturally occurring drifts from live registries and changelogs, including PyPI, npm, GitHub Actions, and Docker Hub. These examples are not generated by the drift model and serve as an external-validity split for environment changes observed in real software ecosystems.

\paragraph{Negative controls.}
False alarms are a first-order benchmark target because a monitor that alerts on every changed URL, package, or configuration value is not useful in a long-lived skill library. We include 49 identity pairs, 300 formatting hard negatives, and 250 semantic hard negatives. Formatting negatives change document form without changing environment assumptions. Semantic hard negatives change actual values---for example commentary versions, alias URLs, homograph values, or non-breaking patches---without violating the skill's operational contract. Together, these 599 controls test whether a method monitors role-bearing obligations rather than raw values.

\begin{wraptable}[18]{r}{7cm}
\vspace{-0.5cm}
\centering
\caption{
\textbf{\dbname{} evaluates both drift detection and false-alarm resistance.}
The benchmark combines controlled drifts, LLM-free real-world drifts, identity pairs, and hard negatives that change text or values without violating operational contracts.
}
\label{tab:dataset}
\small
\renewcommand{\arraystretch}{0.95}\tabcolsep 0.1cm
\begin{tabular}{lcc}
\toprule
 & Controlled & Real-world \\
\midrule
Drift pairs & 174 & 107 \\
Identity no-drift pairs & 49 & --- \\
Formatting hard negatives & 300 & --- \\
Semantic hard negatives & 250 & --- \\
\midrule
Total pairs & 773 & 107 \\
Drift types & 8 & 5 \\
Source skills & 49 & 22 \\
Adjudicated validity & 99.6\% & 100\% \\
Negative controls & \multicolumn{2}{c}{\makecell{599; zero-FP Wilson\\ 95\% CI $[0\%,0.6\%]$}} \\
\bottomrule
\end{tabular}
\end{wraptable}

\paragraph{Human audit.}
To reduce dependence on model-based validation, we additionally conduct a stratified random 20\% human audit of \dbname{} after automatic construction and adjudication. The audit covers 176 of 880 pairs: 35 controlled drifts, 21 real-world drifts, 10 identity pairs, 60 formatting hard negatives, and 50 semantic hard negatives. Human auditors check three properties: whether the drift/no-drift label is correct, whether old and new values are correctly instantiated, and whether hard negatives preserve the skill's operational contracts. In the audited sample, 14/176 pairs require corrections, all involving value normalization, evidence-span adjustment, hard-negative wording, or source metadata; no audited pair changes its drift/no-drift label. All corrections are applied before release. We report the audit protocol and edit categories in \Cref{app:human_audit}. This audit complements cross-family LLM validation: LLM validators provide scalable coverage, while human review directly targets benchmark artifact risk.

\section{\sgname{}: Contract-Based Monitoring}
\label{sec:method}

\begin{figure}[ht]
    \centering
    \includegraphics[width=0.88\linewidth]{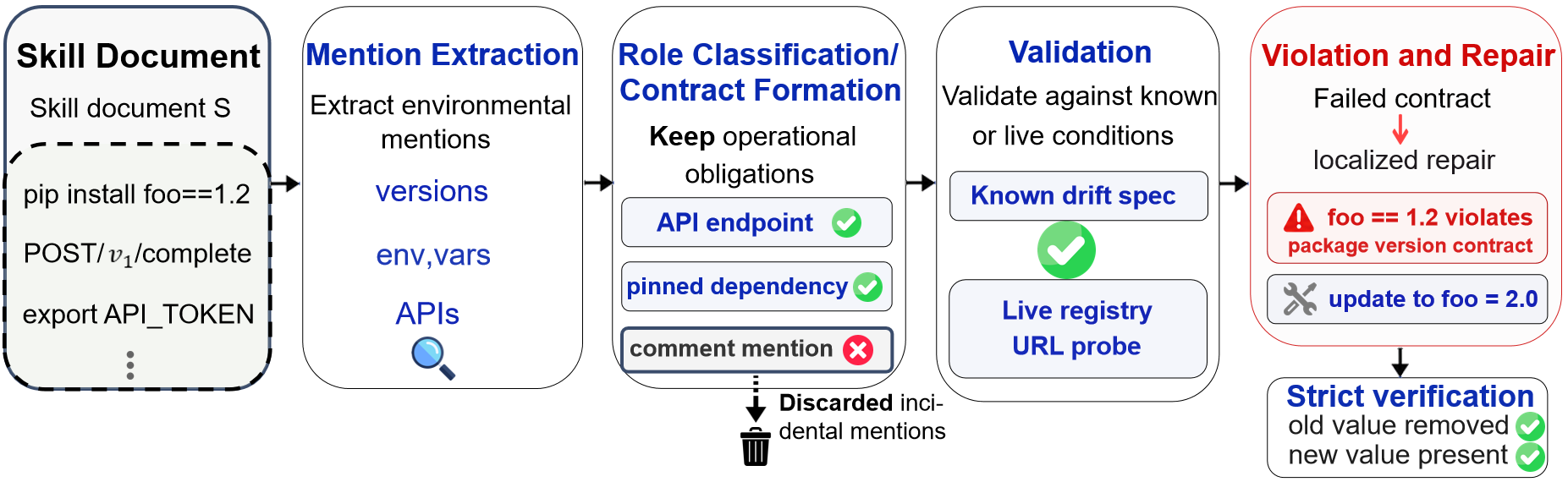}
    \caption{
    \textbf{\sgname{} turns skill maintenance into contract validation.}
    The system first extracts environmental mentions from a skill, keeps only role-bearing operational obligations, validates them against known or live conditions, and uses failed contracts to localize repair.
    The key step is not extraction alone, but separating operational assumptions from incidental mentions before probing the environment.
}
    \label{fig:method_overview}
\end{figure}

\sgname{} is built around one ordering constraint: decide whether an environmental value is role-bearing before probing whether it changed. Contract-free monitors invert this order. They extract URLs, versions, paths, and configuration names, probe them directly, and then treat failed probes as alerts. This creates false alarms whenever an incidental mention changes. \sgname{} instead converts a skill document into operational environment contracts and validates only those contracts against known or live conditions. \Cref{fig:method_overview} summarizes this contract lifecycle.

\paragraph{Problem formulation.}
Let $\skill$ be a skill document and let $\mathcal{M}(\skill)$ denote its environmental mentions: spans that refer to packages, services, URLs, API paths, configuration names, credentials, schemas, or other external assumptions. Only some mentions are operational obligations. \sgname{} extracts an environment contract
\[
\contract(\skill)=\{c_i=(t_i,r_i,v_i,e_i)\}_{i=1}^{n},
\]
where $t_i$ is the contract type, $r_i\in\{\textsc{operational},\textsc{incidental}\}$ is the role of the value in the skill, $v_i$ is the value to validate, and $e_i$ is the evidence span in $\skill$. A known drift event is
\[
d_j=(t_j,o_j,n_j,s_j),
\]
where $t_j$ is the drift type, $o_j$ and $n_j$ are the old and new values, and $s_j$ is the evidence source. A violation is reported only when a drift event matches a contract with $r_i=\textsc{operational}$; matches to incidental mentions are suppressed. This definition is the mechanism behind \sgname{}'s precision-first behavior: it changes the monitored unit from a surface value to a role-bearing obligation.

\subsection{Extracting Role-Bearing Contracts}

The extraction stage solves two different subproblems: finding candidate environment mentions and deciding which of them are operational obligations. Separating these subproblems is necessary because high-recall mention extraction alone is exactly what causes contract-free probes to over-alert.

\paragraph{Candidate mention extraction.}
A deterministic pass identifies syntactically explicit mentions across 15 pattern families, including URLs, version constraints, imports, API paths, authentication patterns, Docker images, GitHub Actions, environment variables, cloud regions, CLI flags, and configuration filenames. This pass is inexpensive and improves coverage, but it does not decide whether a mention should be monitored. Removing broad mention extraction lowers recall; in our sensitivity study, expanding the regex families from 7 to 15 increases regex-only recall on real-world drifts from 45\% to 78\% while preserving 0\% FPR.

\paragraph{Role decision and contract formation.}
A semantic pass~\cite{wei2022chain} converts candidate and implicit assumptions into structured contract records. For each record, the model emits $(t_i,r_i,v_i,e_i)$, where $r_i$ marks whether the value is operational or incidental. This step captures assumptions that are difficult to express as regular expressions, such as expected response formats, tool availability, schema fields, and configuration semantics. It is also the step that prevents extraction from becoming noisy monitoring: without role filtering, CI-style probes produce 40\% false positives, whereas \sgname{} raises zero false alarms across 599 negative controls. Records from both passes are merged and deduplicated; in our implementation, skills typically yield 15--40 contract records, with median 22 for Qwen3.6-Plus. We audit this OPERATIONAL/INCIDENTAL decision independently from downstream drift detection in \Cref{app:role_audit}.

\subsection{Validating Contracts}

Validation asks whether an operational contract is invalidated by known or live environmental evidence. It is deliberately applied after role filtering.

\paragraph{Contract--drift matching.}
After role filtering, \sgname{} matches an operational contract $c_i=(t_i,r_i,v_i,e_i)$ to a drift event $d_j=(t_j,o_j,n_j,s_j)$ using a type-gated conservative predicate:

\[
\operatorname{match}(c_i,d_j)=
\mathbb{I}\!\left[\begin{aligned}
r_i=\textsc{operational}
\land
\operatorname{compatible}(t_i,t_j)
\land
\\\Big(
\operatorname{exact}(\bar v_i,\bar o_j)
\lor
\operatorname{tok}_{2}(\bar v_i,\bar o_j)
\lor
\operatorname{typed}_{1}(t_i,\bar v_i,\bar o_j)
\Big)
\end{aligned}\right],
\]
where $\bar v_i$ and $\bar o_j$ are normalized forms of the contract value and old drift value. The predicate first checks exact or substring overlap, then requires at least two shared meaningful tokens, and finally allows a type-compatible one-token backoff. A violation is reported only when this predicate holds for a contract already labeled \textsc{operational}. Thus, the matcher improves recall after the role decision; it is not a contract-free probe over all mentions in $\mathcal{M}(\skill)$.

\paragraph{Known-drift and live validation.}

In known-drift evaluation, \sgname{} applies $\operatorname{match}$ to each contract--drift pair in $\contract(\skill)\times\drift$. In live monitoring, no old-to-new drift specification is available, so \sgname{} validates operational contracts against public evidence when available, such as package registries, URL responses, and ecosystem conventions. This live mode is precision-oriented rather than exhaustive: contracts without reliable validators, or drifts requiring schema-aware or behavioral reasoning, may be missed.

\paragraph{Live-condition validation.}
When no old-to-new drift specification is available, \sgname{} validates operational contracts against live sources when such validators exist, such as package registries, URL responses, or ecosystem conventions. This mode enables open-world discovery, but it is not exhaustive: contracts without reliable public validators may be missed. We therefore treat live validation as precision-oriented discovery rather than complete environmental verification.Detailed normalization and backoff rules are given in \Cref{app:matcher}.

\subsection{Localizing and Verifying Repair}

A reported violation becomes a localized repair specification. The repair input contains the original skill, the failed contract record, its evidence span, old and new values when available, and category-aware instructions such as import cascading, configuration replacement, or authentication-flow update. \sgname{} generates two candidates at $\tau\in\{0.0,0.2\}$ and selects the first candidate that passes verification. Localization is necessary for repair quality: replacing it with the vague instruction ``the skill is stale'' reduces one-round repair from 78\% to 10\%.

For value-substitution drifts, a repaired skill $\skill'$ passes the literal verifier iff every required new value $n_j$ appears in $\skill'$ and the corresponding old value $o_j$ no longer appears. This verifier prevents false repair success for literal drift updates, but it is not a proof of semantic correctness. It can reject valid repairs that satisfy the new contract without using the exact literal value; we analyze this boundary in \Cref{sec:analysis}.

\subsection{Complexity and Scope}

The deterministic pass is linear in the length of $\skill$. The semantic extraction pass runs once per skill at ingestion or refresh time. Known-drift validation costs $O(|\contract||\drift|)$ lightweight string/type comparisons, and repair is invoked only for reported violations. \sgname{} is not a full verifier for skill correctness. Its scope is narrower: within extraction coverage and available validators, it detects environment-contract violations while suppressing alerts on incidental mentions. This scope matches the deployment goal of a precision-first maintenance monitor rather than an exhaustive execution oracle.

\section{Experiments}
\label{sec:experiments}

Our evaluation asks whether treating skill drift as contract violation changes
the maintenance problem in practice. We organize the evidence around three
questions. First, do role-bearing contracts suppress false alarms without
collapsing drift recall? Second, can the same contract view surface stale skills
under live conditions, where no old-to-new drift specification is provided?
Third, are failed contracts useful repair evidence, rather than only detection
signals? Fine-grained role-label audits, extraction sensitivity, verifier
behavior, and per-type failures are analyzed in \Cref{sec:analysis} and the
appendix.

\paragraph{Setup.}
We evaluate five backbone LLMs from four families: Qwen3.6-Plus and
Qwen3-235B-A22B, DeepSeek-V3.2 and DeepSeek-R1, and GLM-5.1. Detection
baselines include Grep/diff, CI probes without contracts, Dependabot-style
scanning~\cite{fruntke2025automatically}, NL2Contract-style single-pass
extraction~\cite{lahiri2026intent}, and simulated canary execution. Repair
baselines include no localization, plain drift text, Self-refine~\cite{madaan2023self}
with three rounds, and rewriting with full drift specifications. We report
precision, recall, false-positive rate (FPR), strict repair success, bootstrap
95\% confidence intervals with $B{=}10{,}000$, and Fisher exact tests for repair
ablations.

\subsection{Q1. Do contracts change the monitoring error profile?}

The central prediction of the contract view is not merely that a monitor should
find more drift. It is that the unit of monitoring should change: alerts should
be raised for violated operational obligations, not for every changed URL,
version, or configuration value. We therefore evaluate known-drift verification
together with no-drift and hard-negative controls, where many values change but
the skill's operational contract does not.

\begin{wraptable}[13]{r}{7cm}
\vspace{-0.5cm}
\centering
\caption{
\textbf{Known-drift verification across five backbones.}
\sgname{} preserves perfect precision across evaluated backbones, while recall
varies with extraction coverage. This pattern supports the intended
precision-first behavior: role-aware contracts suppress false alarms without
reducing monitoring to trivial abstention.
}
\label{tab:main_results}
\small
\renewcommand{\arraystretch}{0.95}\tabcolsep 0.1cm
\begin{tabular}{lccc}
\toprule
Backbone & Precision & Recall & 95\% CI \\
\midrule
Qwen3.6-Plus & 100\% & 76\% & [62\%, 88\%] \\
DeepSeek-R1 & 100\% & 62\% & [55\%, 70\%] \\
DeepSeek-V3.2 & 100\% & 57\% & [44\%, 70\%] \\
GLM-5.1 & 100\% & 54\% & [40\%, 69\%] \\
Qwen3-235B-A22B & 100\% & 24\% & [17\%, 32\%] \\
\bottomrule
\end{tabular}
\end{wraptable}

Across all five backbones, \sgname{} attains 100\% precision in known-drift
verification; the strongest backbone reaches 76\% recall
(\Cref{tab:main_results}). The same precision behavior holds on negative
controls: over 599 no-drift and hard-negative cases---49 identity pairs, 300
formatting hard negatives, and 250 semantic hard negatives---\sgname{} raises
no false positives (Wilson 95\% CI $[0\%,0.6\%]$). These negatives are not
empty controls: semantic hard negatives change actual values while preserving
the operational contract, so passing them requires the method to distinguish
incidental mentions from role-bearing obligations.

\begin{wrapfigure}[19]{r}{8cm}
\vspace{-0.5cm}
    \centering
    \includegraphics[width=1\linewidth]{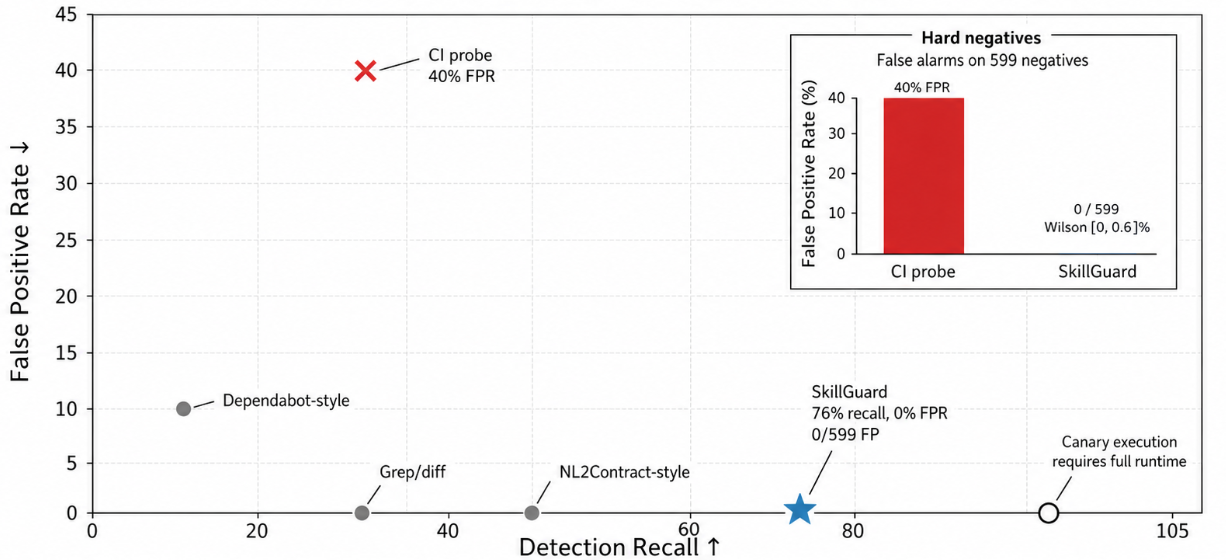}
    \caption{
    \textbf{Contracts change the monitoring error profile.}
    Contract-free probes detect some drift but produce high false-positive
    rates because they probe incidental mentions.
    \sgname{} occupies the precision-first region: 76\% recall and 0\% FPR on
    known drifts, with zero false alarms over 599 no-drift and hard-negative
    cases. Canary execution is an oracle-style upper bound that requires full
    runtime access.
    }
    \label{fig:tradeoff}
\end{wrapfigure}

\Cref{fig:tradeoff} places this result against alternative monitoring
strategies. Grep/diff is precise but low-recall (30\% recall). Dependabot-style
scanning reaches 11\% recall with 10\% FPR, and NL2Contract-style extraction
reaches 45\% recall with 0\% FPR. CI probes without contracts recover some
drift (30\% recall), but incur 40\% FPR because they validate raw environmental
values, including documentation links, examples, and badges. Canary execution
reaches the ideal corner, but only by assuming a fully runnable environment; we
therefore treat it as an upper bound rather than a deployable monitor. The
empirical separation supports the main mechanism: the gain comes from deciding
which assumptions are operational before probing whether they changed.

\subsection{Q2. Does contract monitoring transfer to live discovery?}

Known-drift verification tests whether \sgname{} can match contracts against
explicit drift specifications. A maintenance monitor, however, should also be
useful when those specifications are unavailable. We therefore evaluate
\sgname{} in a pre-registered open-world protocol over 49 real skills. Before
running the monitor, we freeze skill-level drift labels; \sgname{} then extracts
contracts and checks them against live evidence such as registries, URL
responses, and ecosystem conventions.

At the skill level, 22 skills contain at least one confirmed drift and 27 are
fresh. \sgname{} flags 14 skills, yielding 12 true positives, 2 false positives,
10 false negatives, and 25 true negatives. This gives 86\% conservative
precision, 55\% recall, and 7\% FPR. The two apparent false positives were
later adjudicated as genuine drifts, but we keep the pre-adjudication numbers
as the primary result. This choice makes the conclusion deliberately narrow:
\sgname{} can surface stale skills with high precision under live conditions,
but it is not an exhaustive substitute for execution-based checks.

\begin{figure}[ht]
    \centering
    \includegraphics[width=0.7\linewidth]{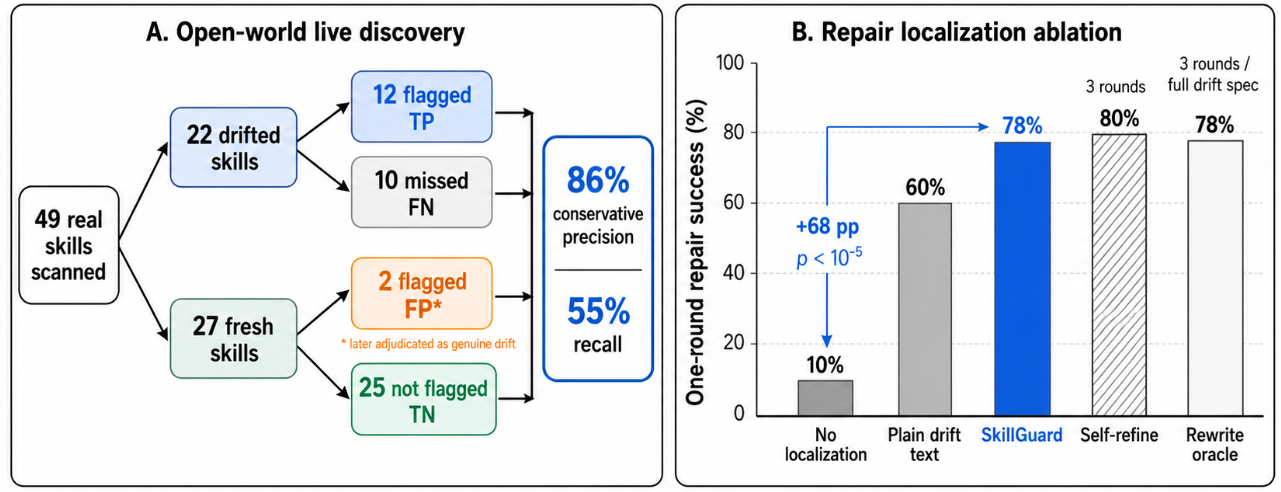}
    \caption{
    \textbf{Contract violations support live maintenance, not just offline detection.}
    (A) In a pre-registered scan of 49 real skills, \sgname{} flags 14 skills
    and achieves 86\% conservative precision and 55\% recall; two apparent
    false positives were later adjudicated as genuine drift.
    (B) Failed contracts localize repair: one-round repair improves from 10\%
    without localization to 78\%, matching stronger multi-round baselines
    without claiming a significant advantage over them.
    }
    \label{fig:live_repair}
\end{figure}

\subsection{Q3. Do failed contracts make repair actionable?}

Detection is useful only if it can guide maintenance. The contract view predicts
that a violation should localize the failed assumption: what value is stale,
where it appears in the skill, and which environmental condition invalidated it.
We test this by comparing one-round contract-guided repair with unlocalized
repair prompts, plain drift text, three-round Self-refine, and rewriting with
full drift specifications. Success is measured by the strict verifier described
in \Cref{sec:method}.

As shown in \Cref{fig:live_repair}(B), telling the model only that ``the skill
is stale'' succeeds in 10\% of cases. Plain drift text reaches 60\%, while
failed contract records reach 78\% in one round ($p<10^{-5}$ versus no
localization). Self-refine reaches 80\%, but uses three rounds; rewriting with
full drift specifications reaches 78\%, but uses stronger oracle information.
Contract-guided repair also uses fewer interactions, averaging roughly 10K
tokens per skill compared with about 18K tokens for three-round Self-refine.

The supported claim is intentionally limited. \sgname{} does not significantly
outperform Self-refine or plain drift text in the current repair sample. Rather,
the result shows that violated contracts are actionable repair evidence: they
close most of the gap between vague repair and stronger multi-round alternatives
while preserving a direct link between the detected failure and the edit request.
The remaining errors---especially authentication changes, schema changes, and
cases rejected by the literal verifier---are analyzed in \Cref{sec:analysis}.

\section{Analysis}
\label{sec:analysis}

The main experiments establish the error profile of \sgname{}: high precision,
useful but incomplete recall, and repair gains from localization. We now ask
what parts of the system produce these behaviors and where the current contract
view reaches its limits.

\subsection{What explains the precision--recall profile?}

\sgname{} is precision-stable but not recall-complete because precision and
recall are controlled by different stages. Precision is mainly a consequence of
the role gate: a changed value is monitored only when it is an
\textsc{operational} obligation of the skill. Recall, in contrast, is bounded by
whether the relevant obligation is extracted and whether a validator can check
it.

A standalone role-label audit supports this decomposition. Two annotators label
200 environmental mentions, and disagreements are adjudicated before evaluation.
Treating \textsc{operational} as the positive class, \sgname{} achieves 89.5\%
precision, 83.4\% recall, and 86.3\% F1 overall. The split-level pattern matches
the end-to-end behavior: on semantic hard negatives, role precision reaches
100.0\%, indicating that the model is conservative on value-changing but
contract-preserving cases; role recall on the same split is 72.7\%, indicating a
source of missed operational assumptions.

Extraction sensitivity points to the other side of the trade-off. Expanding the
deterministic extractor from 7 to 15 pattern families increases regex-only
recall from 13\% to 33\% on the controlled split and from 45\% to 78\% on
real-world drift, while FPR remains 0\%. The added patterns cover environmental
forms common in real skills, including Docker images, GitHub Actions versions,
environment variables, cloud regions, branch names, CLI flags, and
configuration filenames. Thus, the current recall gap is not evidence that
precision requires abstention: coverage can improve without weakening the
role-bearing contract filter.

\subsection{What does repair evidence actually show?}

The repair experiment should be read as a localization result, not as a claim
that \sgname{} dominates all repair methods. A failed contract identifies which
operational assumption was invalidated, where it appears in the skill, and what
evidence triggered the violation. This converts detection into a constrained
edit request rather than an open-ended rewrite.

This localization signal is much stronger than a vague stale-skill prompt:
repair success rises from 10\% without localization to 78\% with failed
contracts in one round. However, the stronger baselines remain competitive:
plain drift text reaches 60\%, and three-round Self-refine reaches 80\%. The
differences between \sgname{} and plain drift text or Self-refine are not
statistically significant in our current sample. The supported conclusion is
therefore narrower and more useful: violated contracts are actionable repair
evidence, closing most of the gap between vague repair and stronger multi-round
alternatives while using fewer interactions.

\subsection{Where does the current contract view break?}

Two boundaries remain. The first is a measurement boundary. Our strict verifier
accepts value-substitution repairs only when the required new value appears and
the stale value disappears. This avoids many false repair successes, but it is
not a semantic proof of skill correctness. A type-aware verifier accepts 168/174
cases, compared with 165/174 for the literal verifier, leaving a roughly 2\%
conservative rejection gap. Most of this gap appears in dependency updates,
where a semantically valid edit may satisfy the new dependency constraint
without preserving the exact literal form.

The second boundary is a mechanism boundary. \sgname{} is strongest when drift
invalidates a localizable environment assumption, such as a URL, version, API
path, or configuration value. It is weaker when the repair requires changing an
interaction mechanism. In the Qwen3.6-Plus per-type breakdown, URL changes and
version bumps are detected at 100\%, while configuration, deprecation, and
schema changes fall to 60--67\%. Authentication drift is the hardest case in
this slice: detection reaches 67\%, but repair succeeds in 0\% of the three
examples. These failures do not contradict the contract framing; they mark the
point where surface contracts should be complemented by schema-aware validators,
execution-backed checks, or richer repair plans.

\section{Conclusion}
\label{sec:conclusion}

Long-lived agent skills do not fail merely because the environment changes;
they fail when those changes invalidate assumptions the skill operationally
depends on. This paper framed skill drift as contract violation and introduced
\sgname{}, which extracts environment contracts, validates only role-bearing
assumptions, and uses failed contracts to localize repair. This granularity
changes the monitoring error profile: contract-free CI probes incur 40\% FPR,
whereas \sgname{} raises zero false alarms over 599 no-drift and hard-negative
cases; in open-world discovery, it surfaces live drift with 86\% conservative
precision. The same violations make repair local, improving one-round repair
from 10\% without localization to 78\%. Together with \dbname{}, these results
make skill degradation measurable as a maintenance problem, while leaving
exhaustive semantic verification to complementary execution-based methods.


\bibliographystyle{plainnat}
\bibliography{references}

\appendix

\section{Appendix Overview}
\label{app:overview}

The appendix provides the evidence needed to audit the claims made in the main
paper. \Cref{app:notation,app:limitations,app:algorithm,app:matcher} specify the contract
representation, limitations, algorithmic ordering, and conservative matching rules used by
\sgname{}. \Cref{app:driftbench_details,app:human_audit} describe the
construction and validation of \dbname{}. \Cref{app:role_audit,app:negative_controls,app:baselines_sensitivity}
support the main precision and recall claims. \Cref{app:openworld,app:repair_verifier,app:failure_taxonomy}
give additional evidence for live discovery, repair localization, verifier
behavior, and failure modes. \Cref{app:implementation} summarizes
implementation and reproducibility details.

\section{Notation and Scope}
\label{app:notation}

\begin{table}[ht]
\centering
\caption{
\textbf{Notation used by \sgname{}.}
The role variable $r_i$ is the key distinction: \textsc{operational} mentions
are monitored as contracts, while \textsc{incidental} mentions are suppressed
before validation.
}
\label{tab:notation}
\small
\begin{tabular}{ll}
\toprule
Symbol & Meaning \\
\midrule
$\skill$ & Skill document \\
$\mathcal{M}(\skill)$ & Environmental mentions in $\skill$ \\
$\contract(\skill)$ & Extracted environment contract set \\
$c_i=(t_i,r_i,v_i,e_i)$ & Contract record: type, role, value, evidence span \\
$r_i$ & Role: \textsc{operational} or \textsc{incidental} \\
$d_j=(t_j,o_j,n_j,s_j)$ & Drift event: type, old value, new value, source \\
$\operatorname{match}(c_i,d_j)$ & Contract--drift matching predicate \\
$\mathcal{V}$ & Reported contract violations \\
$\skill'$ & Repaired skill document \\
\bottomrule
\end{tabular}
\end{table}

\paragraph{Role semantics.}
A mention is \textsc{operational} if the skill relies on the value for execution
or correctness, such as a pinned dependency, API endpoint, schema field,
authentication mechanism, environment variable, or configuration value. A
mention is \textsc{incidental} if changing it does not alter the skill's
operational behavior, such as a comment, tutorial citation, badge, or
documentation example.

\paragraph{Scope.}
\sgname{} detects violations of environment-dependent assumptions. It is not a
complete semantic verifier for skill correctness, an exhaustive live drift
discovery system, or a proof that a repaired skill is correct under all
executions. This scope matches the intended use of \sgname{} as a precision-first
maintenance monitor.

\section{Limitations}
\label{app:limitations}

\sgname{} is a precision-first monitor for environment-contract violations, not
a complete verifier of skill correctness. Its guarantees therefore apply only
within the coverage of extracted contracts and available validators. Assumptions
that are implicit, behavioral, or observable only through execution can be
missed. The role-label audit shows that the
\textsc{operational}/\textsc{incidental} distinction is reliable but imperfect;
missed operational roles remain one source of false negatives. Similarly, live
validators based on package registries, URL responses, and ecosystem
conventions provide useful discovery signals, but they do not replace
execution-based canaries when exhaustive coverage is required.

Repair has the same boundary. Failed contracts localize stale assumptions, but
they do not by themselves solve mechanism-level rewrites such as authentication
flows, coupled schema changes, or behavioral API migrations. The strict repair
verifier is also an acceptance criterion for value substitutions, not a semantic
correctness proof; it can reject valid edits that satisfy the updated contract
without preserving the exact literal form. Finally, our open-world study covers
49 real skills. Larger continuous deployments are needed to measure long-term
alert fatigue, human acceptance of suggested repairs, and how contract monitors
should be combined with execution-backed checks.

\section{Algorithmic Details}
\label{app:algorithm}

\begin{algorithm}[ht]
\caption{\sgname{} contract-based monitoring}
\label{alg:skillguard}
\small
\begin{algorithmic}[1]
\Require skill document $\skill$, optional drift events $\drift$
\State $\mathcal{M} \gets$ ExtractMentions$(\skill)$
\State $\contract \gets$ FormContracts$(\skill,\mathcal{M})$
\State $\mathcal{V}\gets \emptyset$
\For{$c_i=(t_i,r_i,v_i,e_i)\in \contract$}
    \If{$r_i=\textsc{incidental}$}
        \State continue
    \EndIf
    \State $\mathcal{D}_i \gets$ KnownDriftOrLiveEvidence$(c_i,\drift)$
    \For{$d_j\in \mathcal{D}_i$}
        \If{$\operatorname{match}(c_i,d_j)$}
            \State $\mathcal{V}\gets \mathcal{V}\cup\{(c_i,d_j)\}$
        \EndIf
    \EndFor
\EndFor
\State $\skill' \gets$ RepairWithViolations$(\skill,\mathcal{V})$
\State \Return violations $\mathcal{V}$, optional verified repair $\skill'$
\end{algorithmic}
\end{algorithm}

\paragraph{Design rationale.}
The critical ordering in \Cref{alg:skillguard} is that role filtering precedes
validation. Contract-free probes validate every extracted URL, version, path, or
configuration value and therefore over-alert on incidental mentions. \sgname{}
first decides whether the mention is an operational obligation, and only then
validates it against known drift specifications or live evidence. This ordering
is the algorithmic form of the paper's central claim: skill drift is a
role-bearing contract violation, not a raw value change.

\section{Matcher Normalization and Backoff Rules}
\label{app:matcher}

\paragraph{Normalization.}
Before matching, \sgname{} normalizes contract values and drift values by
lowercasing package and API identifiers, stripping surrounding punctuation,
normalizing URL slashes, splitting common separators
(\texttt{/}, \texttt{-}, \texttt{\_}, \texttt{.}), and removing tokens such as
protocol prefixes or file extensions when they do not determine the contract
type. Normalization is type-aware: version strings, package names, API paths,
environment variables, and URLs use separate tokenization rules.

\begin{table}[ht]
\centering
\caption{
\textbf{Conservative matching after role filtering.}
The matcher can improve recall only after the role gate has removed incidental
mentions; otherwise the same normalization would create false alarms on comments,
badges, and examples.
}
\label{tab:matcher_rules}
\small
\begin{tabular}{llp{0.46\linewidth}}
\toprule
Level & Rule & Purpose \\
\midrule
Role gate & $r_i=\textsc{operational}$ &
Suppress comments, badges, tutorials, and documentation-only mentions before
validation. \\
Type gate & $\operatorname{compatible}(t_i,t_j)$ &
Prevent near-collisions across unrelated contract types. \\
Exact & $\operatorname{exact}(\bar v_i,\bar o_j)$ &
Capture literal substitutions and normalized substring matches. \\
Token-2 & $\operatorname{tok}_{2}(\bar v_i,\bar o_j)$ &
Recover drift when formatting or path separators change. \\
Typed-1 & $\operatorname{typed}_{1}(t_i,\bar v_i,\bar o_j)$ &
Allow conservative one-token backoff only within compatible contract types. \\
Reject & otherwise &
Avoid alerting on raw environmental values without operational role support. \\
\bottomrule
\end{tabular}
\end{table}

\paragraph{Why not use a stronger semantic matcher?}
The matcher is intentionally conservative because \sgname{} is designed as a
precision-first monitor. More aggressive semantic normalization could recover
schema renames, authentication-flow migrations, or API behavior changes, but it
could also create near-collisions among incidental examples. We therefore use
conservative surface matching in the main system and treat schema-aware or
execution-backed validators as complementary extensions for contract types with
weak surface evidence.

\section{\dbname{} Construction and Validation}
\label{app:driftbench_details}

\begin{table}[ht]
\centering
\caption{
\textbf{\dbname{} evaluates both drift recall and false-alarm resistance.}
The benchmark combines controlled drifts, LLM-free real-world drifts, identity
pairs, and hard negatives that change text or values without violating
operational contracts.
}
\label{tab:appendix_dataset}
\small
\begin{tabular}{lrr}
\toprule
 & Controlled & Real-world \\
\midrule
Drift pairs & 174 & 107 \\
Identity no-drift pairs & 49 & --- \\
Formatting hard negatives & 300 & --- \\
Semantic hard negatives & 250 & --- \\
\midrule
Total pairs & 773 & 107 \\
Drift types & 8 & 5 \\
Source skills & 49 & 22 \\
Adjudicated validity & 99.6\% & 100\% \\
Negative controls & \multicolumn{2}{c}{599; zero-FP Wilson 95\% CI $[0\%,0.6\%]$} \\
\bottomrule
\end{tabular}
\end{table}

\paragraph{Controlled drift generation.}
The controlled split starts from 49 agent skills and covers eight drift
categories: URL change, version bump, configuration change, API migration,
deprecation, schema change, authentication change, and dependency update.
Qwen3.6-Plus proposes 6--8 drift scenarios per skill, yielding 422 candidate
pairs. Validated candidates form the 174-pair controlled drift split used in the
main evaluation.

\paragraph{Cross-family validation and adjudication.}
To reduce generator artifacts, candidate pairs are checked using cross-family
validation and registry-backed adjudication. DeepSeek-R1 independently judges
candidate validity. Disputed cases are checked against registry or changelog
evidence and then reviewed by the authors. The final adjudicated validity is
99.6\% for the controlled split.

\paragraph{LLM-free real-world split.}
The real-world split contains 107 naturally occurring drifts collected from live
registries and changelogs, including package registries, repository metadata,
and ecosystem documentation. No LLM is used to generate these drift instances.
This split is intended to test whether contract monitoring transfers beyond
model-generated drift pairs.

\section{Human Audit of \dbname{}}
\label{app:human_audit}

\paragraph{Purpose.}
The human audit tests benchmark validity independently of the model-based
construction and adjudication pipeline. It targets three possible artifact
sources: incorrect drift/no-drift labels, incorrect old/new value instantiation,
and hard negatives that accidentally violate an operational contract.

\paragraph{Sampling.}
We use stratified random sampling over the five benchmark splits. The audit
covers 176 of 880 pairs (20\%): 35 controlled drift pairs, 21 real-world drift
pairs, 10 identity no-drift pairs, 60 formatting hard negatives, and 50 semantic
hard negatives. Stratification ensures that the audit covers both positive drift
examples and the negative controls that support the false-alarm claim.

\begin{table}[ht]
\centering
\caption{
\textbf{Random 20\% human audit of \dbname{}.}
Corrections affect value rendering, spans, wording, or metadata; no audited
example changes its drift/no-drift label.
}
\label{tab:human_audit}
\small
\begin{tabular}{lrrrr}
\toprule
Split & Total & Audited & Corrected & Label flips \\
\midrule
Controlled drift pairs & 174 & 35 & 4 & 0 \\
Real-world drift pairs & 107 & 21 & 2 & 0 \\
Identity no-drift pairs & 49 & 10 & 0 & 0 \\
Formatting hard negatives & 300 & 60 & 3 & 0 \\
Semantic hard negatives & 250 & 50 & 5 & 0 \\
\midrule
Total & 880 & 176 & 14 & 0 \\
\bottomrule
\end{tabular}
\end{table}

\begin{table}[ht]
\centering
\caption{
\textbf{Human-audit correction categories.}
The audit improves released pair quality without changing any sampled
drift/no-drift label.
}
\label{tab:human_audit_edits}
\small
\begin{tabular}{lrp{0.55\linewidth}}
\toprule
Correction type & Count & Description \\
\midrule
Value normalization & 5 &
Standardize package names, version strings, endpoint formatting, or old/new
value rendering. \\
Evidence-span adjustment & 4 &
Move the evidence span to the exact line or snippet supporting the contract or
drift. \\
Hard-negative wording & 3 &
Clarify transformations so that they change surface form or incidental values
without violating an operational contract. \\
Source metadata & 2 &
Update registry, changelog, or URL metadata without changing the pair label. \\
\midrule
Total & 14 & --- \\
\bottomrule
\end{tabular}
\end{table}

\paragraph{Interpretation.}
The audit supports \dbname{} as a benchmark for skill degradation rather than
raw text change detection. All corrected cases are retained after editing, and
no audited pair changes its drift/no-drift label. The audit is distinct from the
role-label audit in \Cref{app:role_audit}: this section verifies the benchmark
pairs, while the next section evaluates the OPERATIONAL/INCIDENTAL decision
used by \sgname{}.

\section{Standalone Role-Label Audit}
\label{app:role_audit}

\paragraph{Purpose.}
The OPERATIONAL/INCIDENTAL decision is the central intermediate variable in
\sgname{}. The end-to-end false-positive result shows that this decision is
useful for monitoring, but it does not by itself measure role-label accuracy. We
therefore perform a standalone audit of role classification on human-labeled
environmental mentions.

\paragraph{Protocol.}
Two annotators independently label 200 environmental mentions drawn from
controlled skills, semantic hard negatives, and open-world skills. Annotators
mark a mention as \textsc{operational} if the skill relies on it for execution
or correctness, and \textsc{incidental} otherwise. Disagreements are adjudicated
before evaluation. We treat \textsc{operational} as the positive class and
compare \sgname{}'s role predictions against the adjudicated human labels.

\begin{table}[ht]
  \centering
  \caption{
  \textbf{Standalone role-label audit.}
  Positive denotes \textsc{operational}. This audit evaluates the
  OPERATIONAL/INCIDENTAL decision independently from end-to-end drift detection
  using adjudicated human labels.
  }
  \label{tab:role_audit}
  \small
  \begin{tabular}{lrrrr}
  \toprule
  Split & Mentions & Precision & Recall & F1 \\
  \midrule
  Controlled skills       & 43  & 81.2\%  & 89.7\% & 85.2\% \\
  Semantic hard negatives & 44  & 100.0\% & 72.7\% & 84.2\% \\
  Open-world skills       & 113 & 88.6\%  & 86.7\% & 87.6\% \\
  \midrule
  \textbf{All}            & \textbf{200} & \textbf{89.5\%} & \textbf{83.4\%} & \textbf{86.3\%} \\
  \bottomrule
  \end{tabular}
\end{table}

\paragraph{Interpretation.}
The audit supports the mechanism behind \sgname{}'s precision-first behavior.
High role precision means that predicted operational mentions are usually true
operational assumptions. This is especially important on semantic hard
negatives, where values change but contracts should remain valid. Lower recall
on the same split indicates a source of missed operational assumptions, which
matches the main paper's conclusion that recall is bounded by extraction and
role-label coverage. The audit therefore strengthens the central claim without
implying exhaustive drift detection.

\newpage

\section{Additional Detection Results}
\label{app:model_results}

\begin{table}[ht]
\centering
\caption{
\textbf{Model-level known-drift verification.}
Precision is stable across evaluated backbones, while recall varies with
extraction and model behavior.
}
\label{tab:appendix_model_results}
\small
\begin{tabular}{llrrc}
\toprule
Model & Family & Precision & Recall & 95\% CI \\
\midrule
Qwen3.6-Plus & Qwen & 100\% & 76\% & [62\%,88\%] \\
DeepSeek-R1 & DeepSeek & 100\% & 62\% & [55\%,70\%] \\
DeepSeek-V3.2 & DeepSeek & 100\% & 57\% & [44\%,70\%] \\
GLM-5.1 & GLM & 100\% & 54\% & [40\%,69\%] \\
Qwen3-235B-A22B & Qwen MoE & 100\% & 24\% & [17\%,32\%] \\
\bottomrule
\end{tabular}
\end{table}

\paragraph{Interpretation.}
We use model-level results to characterize the precision--recall profile of
\sgname{}, not to rank backbone capability. The stable precision across models
supports the role-filtering mechanism, while recall variation indicates
coverage and extraction sensitivity.

\section{Negative Controls and Statistical Bounds}
\label{app:negative_controls}

\paragraph{Formatting hard negatives.}
The 300 formatting hard negatives apply deterministic surface edits such as
section reordering, whitespace changes, markdown normalization, and harmless
structural edits. They test whether a detector overreacts to document changes
that do not alter environment assumptions.

\paragraph{Semantic hard negatives.}
The 250 semantic hard negatives change actual values while preserving
operational contracts. Examples include commentary versions that are not
dependencies, alias URLs, homograph values, and non-breaking patches. These
cases directly test the core role distinction: a value can change without
becoming a skill-relevant violation.

\begin{table}[ht]
\centering
\caption{
\textbf{False-positive bounds on no-drift and hard-negative cases.}
The combined 599 negative controls provide the statistical support for the
paper's false-alarm claim.
}
\label{tab:appendix_fpr}
\small
\begin{tabular}{lrrcc}
\toprule
Negative split & FP & Total & Wilson 95\% CI & Clopper--Pearson 95\% CI \\
\midrule
Identity no-drift & 0 & 49 & [0.0\%, 7.3\%] & [0.0\%, 7.3\%] \\
Formatting hard negatives & 0 & 300 & [0.0\%, 1.3\%] & [0.0\%, 1.2\%] \\
Semantic hard negatives & 0 & 250 & [0.0\%, 1.5\%] & [0.0\%, 1.5\%] \\
\midrule
All negatives & 0 & 599 & [0.0\%, 0.6\%] & [0.0\%, 0.6\%] \\
\bottomrule
\end{tabular}
\end{table}

\section{Detection Baselines and Extraction Sensitivity}
\label{app:baselines_sensitivity}

\begin{table}[ht]
\centering
\caption{
\textbf{Detection baselines.}
Contract-free probes can detect some drift but incur false alarms because they
validate raw values rather than role-bearing obligations.
}
\label{tab:appendix_baselines}
\small
\begin{tabular}{lrrrl}
\toprule
Method & Precision & Recall & FPR & Requirement \\
\midrule
Grep/diff & 100\% & 30\% & 0\% & old+new text \\
CI probes without contracts & 60\% & 30\% & 40\% & live environment \\
Dependabot-style scanning & 80\% & 11\% & 10\% & version database \\
NL2Contract-style extraction & 100\% & 45\% & 0\% & single LLM pass \\
Static config checker & 0\% & 0\% & 0\% & rule library \\
\sgname{} & 100\% & 76\% & 0\% & contracts + drift evidence \\
Canary execution & 100\% & 100\% & 0\% & full runtime \\
\bottomrule
\end{tabular}
\end{table}

\begin{table}[ht]
\centering
\caption{
\textbf{Extraction sensitivity.}
Expanding deterministic mention coverage improves recall while preserving
zero-FPR behavior, suggesting that recall is coverage-limited rather than
precision-limited.
}
\label{tab:appendix_regex}
\small
\begin{tabular}{lrrr}
\toprule
Metric & 7-family regex & 15-family regex & Change \\
\midrule
Controlled-split regex recall & 13\% & 33\% & +20 pp \\
Real-world regex recall & 45\% & 78\% & +33 pp \\
Mean contracts per skill & 3.4 & 11.7 & +8.3 \\
FPR & 0\% & 0\% & no change \\
\bottomrule
\end{tabular}
\end{table}

\paragraph{Additional pattern families.}
The expanded extractor adds Docker image versions, npm \texttt{@version}
patterns, GitHub Actions versions, environment variables, cloud regions, git
branch names, CLI flags, and configuration filenames. These patterns explain
much of the recall gain on real-world drift.

\paragraph{External untailored set.}
On a small external set outside the original categories, the original pipeline
detects 0/7 drifts, while the improved pipeline detects 7/7 drifts with 0 false
positives. We treat this as supportive evidence for improved coverage, not as a
decisive generalization claim because the set is small.

\section{Open-World Discovery Details}
\label{app:openworld}

\begin{table}[ht]
\centering
\caption{
\textbf{Open-world discovery at the skill level.}
We report conservative pre-adjudication values as primary, even though the two
apparent false positives were later judged genuine drift.
}
\label{tab:appendix_openworld}
\small
\begin{tabular}{lr}
\toprule
Quantity & Value \\
\midrule
Skills scanned & 49 \\
Skills with confirmed drift & 22 \\
Fresh skills & 27 \\
Skills flagged by \sgname{} & 14 \\
True positives & 12 \\
False positives, pre-adjudication & 2 \\
False negatives & 10 \\
True negatives & 25 \\
Conservative precision & 86\% \\
Recall & 55\% \\
FPR & 7\% \\
\bottomrule
\end{tabular}
\end{table}

\paragraph{Flagged skills.}
The 14 flagged skills are \texttt{analyze-ci},
\texttt{bazel-build-optimization}, \texttt{changelog-automation},
\texttt{dbt-transformation-patterns}, \texttt{distributed-tracing},
\texttt{gitops-workflow}, \texttt{llm-evaluation},
\texttt{nx-workspace-patterns}, \texttt{python-packaging},
\texttt{rag-implementation}, \texttt{security-review},
\texttt{similarity-search-patterns}, \texttt{spark-optimization}, and
\texttt{vector-index-tuning}.

\paragraph{Blinded adjudication.}
The two apparent false positives were re-evaluated after the primary protocol.
A \texttt{dbt} documentation redirect returned a 301 and was judged genuine
drift with high confidence. A broken MLflow URL returned a 404 and was judged
genuine drift with medium confidence. We nevertheless report the
pre-adjudication precision in the main paper to avoid overstating
live-monitoring performance.

\paragraph{Live-monitoring simulation.}
A separate live-monitoring simulation over 49 skills detects 8 of 11 drifted
skills, with 0 false positives among 38 fresh skills. The repair stage succeeds
on 3 detected skills, giving 38\% conditional repair and 27\% end-to-end
success. We include this as deployment-oriented evidence but do not use it as
the primary open-world result because the pre-registered discovery study is the
cleaner evaluation.

\section{Repair Ablations, Cost, and Verifier Behavior}
\label{app:repair_verifier}

\begin{table}[ht]
\centering
\caption{
\textbf{Repair guidance ablation.}
Failed contracts substantially improve over no localization, but do not
significantly outperform plain drift text or three-round Self-refine in the
current sample.
}
\label{tab:appendix_repair}
\small
\begin{tabular}{lrrrl}
\toprule
Repair signal & Success & $n$ & Rounds & $p$ vs.\ \sgname{} \\
\midrule
No localization & 2/20 = 10\% & 20 & 1 & $p<10^{-5}$ \\
Plain drift text & 12/20 = 60\% & 20 & 1 & $p=0.213$ \\
Self-refine & 16/20 = 80\% & 20 & 3 & $p=1.000$ \\
\sgname{} failed contracts & 25/32 = 78\% & 32 & 1 & --- \\
Rewrite with full drift spec & 25/32 = 78\% & 32 & 3 & $p=1.000$ \\
\bottomrule
\end{tabular}
\end{table}

\begin{table}[ht]
\centering
\caption{
\textbf{Repair cost.}
Contract-guided repair reaches comparable success to multi-round alternatives
with fewer rounds and lower token use.
}
\label{tab:appendix_cost}
\small
\begin{tabular}{lrrr}
\toprule
Method & Tokens per skill & Rounds & Repair success \\
\midrule
\sgname{} failed contracts & $\sim$10K & 1 & 78\% \\
Self-refine & $\sim$18K & 3 & 80\% \\
Rewrite with full drift spec & $\sim$12K & 3 & 78\% \\
\bottomrule
\end{tabular}
\end{table}

\begin{table}[ht]
\centering
\caption{
\textbf{Literal versus type-aware repair verification.}
The literal verifier is conservative: it rejects a small number of repairs that
a type-aware verifier accepts, mainly for dependency updates.
}
\label{tab:appendix_verifier}
\small
\begin{tabular}{lrrr}
\toprule
Drift type & Literal pass & Type-aware pass & Gap \\
\midrule
API migration & 100\% & 100\% & 0 pp \\
Authentication change & 79\% & 79\% & 0 pp \\
Configuration change & 96\% & 96\% & 0 pp \\
Dependency update & 82\% & 100\% & +18 pp \\
Deprecation & 100\% & 100\% & 0 pp \\
Schema change & 95\% & 95\% & 0 pp \\
URL change & 100\% & 100\% & 0 pp \\
Version bump & 100\% & 100\% & 0 pp \\
\midrule
Overall & 165/174 = 95\% & 168/174 = 97\% & +2 pp \\
\bottomrule
\end{tabular}
\end{table}

\paragraph{Interpretation.}
The repair evidence supports a localization claim, not a dominance claim. Failed
contracts are much better than vague repair instructions, but the current sample
does not show a statistically significant advantage over plain drift text or
Self-refine. The verifier analysis also explains why repair scores should be
read conservatively: literal checking avoids false repair success but is not a
semantic correctness proof.

\newpage
\section{Per-Type Failure Taxonomy}
\label{app:failure_taxonomy}

\begin{table}[ht]
\centering
\caption{
\textbf{Per-drift-type detection and repair for Qwen3.6-Plus.}
Small per-type counts make these results diagnostic rather than definitive; they
identify where richer validators and repair plans are needed.
}
\label{tab:appendix_pertype}
\small
\begin{tabular}{lrrrr}
\toprule
Drift type & $n$ & Detection & SG repair & Rewrite repair \\
\midrule
URL change & 4 & 100\% & 100\% & 75\% \\
Version bump & 4 & 100\% & 100\% & 100\% \\
API migration & 6 & 83\% & 83\% & 100\% \\
Dependency update & 5 & 80\% & 40\% & 80\% \\
Configuration change & 9 & 67\% & 56\% & 89\% \\
Deprecation & 6 & 67\% & 50\% & 67\% \\
Authentication change & 3 & 67\% & 0\% & 0\% \\
Schema change & 5 & 60\% & 40\% & 80\% \\
\bottomrule
\end{tabular}
\end{table}

\paragraph{Failure modes.}
Authentication drift often requires replacing an interaction mechanism rather
than substituting a value. Schema changes are frequently embedded in prose or
example outputs, making extraction harder. Behavioral coupling arises when
multiple dependent fields must change together. These failures reflect limits
in extraction and repair localization, not evidence that role-bearing validation
itself causes false alarms.

\section{Implementation and Reproducibility Details}
\label{app:implementation}

\paragraph{Extraction.}
The deterministic pass uses 15 regex families to identify explicit
environmental mentions. The semantic pass performs role classification and emits
contract records $(t_i,r_i,v_i,e_i)$. The two passes are merged and deduplicated
before validation.

\paragraph{Validation.}
Known-drift validation uses exact substring matching, token-overlap matching,
and type-compatible backoff matching. Live validation uses available public
evidence such as package registries, URL responses, and ecosystem conventions.
Contracts without reliable validators may be missed, which is why live
monitoring is treated as precision-oriented discovery rather than exhaustive
verification.

\paragraph{Repair.}
For each reported violation, the repair prompt includes the original skill, the
failed contract, the evidence span, old and new values when available, and
category-aware repair instructions. The system generates candidates at
temperatures $\tau\in\{0.0,0.2\}$ and accepts the first candidate that passes
verification.

\paragraph{Statistical protocol.}
Detection confidence intervals are bootstrap 95\% intervals with
$B=10{,}000$ resamples unless otherwise specified. Zero-false-positive claims
are reported with Wilson and Clopper--Pearson intervals. Repair ablation
significance uses Fisher exact tests. We report non-significant comparisons
explicitly rather than converting them into positive claims.

\paragraph{Reproducibility boundary.}
All reported numbers are derived from JSON result files generated by the
experiment scripts. We use model-level results to characterize the
precision--recall profile of \sgname{} across backbone families, and report
implementation details, prompts, and statistical protocols to support
reproducibility.

\paragraph{Reproducibility package.}
The anonymized supplementary package contains the benchmark pairs, processed
contract records, result JSON files, prompts used for extraction/validation/repair,
and scripts for reproducing the tables and figures in the paper. In particular,
\texttt{scripts/build\_driftbench.py} constructs \dbname{} from the released
source skills and drift specifications, \texttt{scripts/run\_detection.py}
reproduces the known-drift and negative-control evaluations,
\texttt{scripts/run\_openworld.py} reproduces the live-discovery protocol, and
\texttt{scripts/run\_repair.py} reproduces the repair ablations. The exact API
model names, decoding temperatures, and statistical-test scripts are included
in the supplementary material. Closed-source hosted models are accessed through
their public APIs; reproducing model-dependent numbers therefore requires access
to the corresponding model endpoints.

\section{Broader Impacts and Safeguards}
\label{app:broader_impacts}

\paragraph{Positive impacts.}
Long-lived agent skill libraries are increasingly used to automate software
maintenance, testing, deployment, and tool interaction. A precision-first drift
monitor can reduce stale automation, lower alert fatigue, and make maintenance
failures easier to diagnose by localizing violated environment assumptions.

\paragraph{Risks and mitigations.}
The same capability could make agent systems more robust in settings where the
underlying tasks are undesirable, or surface stale dependencies that reveal
maintenance weaknesses. We mitigate these risks by releasing a benchmark of
skill degradation rather than credentials, exploits, or private infrastructure
states; by excluding secrets and private user data from the benchmark; and by
framing \sgname{} as a maintenance monitor rather than an autonomous deployment
or exploitation system. Repair suggestions should be reviewed before being
applied in production, especially for authentication, schema, or behaviorally
coupled changes.

\section{Assets and Licensing}
\label{app:assets_licensing}

\paragraph{Third-party assets and hosted services.}
Table~\ref{tab:assets_licenses} summarizes the external assets, hosted model
APIs, public metadata sources, and newly introduced artifacts used in this work.
For hosted LLMs, we use provider APIs only and do not redistribute model weights,
provider code, or proprietary service artifacts. API access is therefore governed
by the corresponding provider terms rather than by an open-source model license.
For public registry and platform metadata, we only record publicly visible drift
evidence, such as package versions, release records, repository metadata,
workflow metadata, and container metadata, together with access dates. We do not
redistribute third-party package source code, container images, private
repository contents, or provider-side logs. For third-party skill and benchmark
assets, we pin the exact source version used in our experiments and preserve
upstream copyright notices and nested license files.

\paragraph{Release of \dbname{}.}
We release the artifacts introduced by this paper under a permissive split
license: benchmark data, annotations, contract labels, evaluation metadata, and
documentation are released under the Creative Commons Attribution 4.0
International License (CC BY 4.0), while evaluation code and utility scripts are
released under the MIT License. This license applies only to artifacts authored
by us. Any third-party assets referenced, derived from, or bundled for
reproducibility remain subject to their original licenses or platform terms. For
the anonymous submission, we provide an anonymized reviewer-accessible artifact
with the same license files, asset inventory, and pinned versions. Upon
acceptance, we will de-anonymize the repository, preserve the same license
structure, and archive a versioned release.

\begin{table}[t]
\centering
\caption{
\textbf{Assets, licenses, and provider terms.}
Hosted LLMs are accessed through provider APIs; no model weights are redistributed.
Public registry and platform metadata are used only as drift evidence and are
recorded with access dates. Third-party assets remain under their original
licenses or platform terms.
}
\label{tab:assets_licenses}
\small
\begin{tabular}{p{0.23\linewidth}p{0.25\linewidth}p{0.24\linewidth}p{0.20\linewidth}}
\toprule
Asset / Service & Use in this work & Version / Access record & License / Terms \\
\midrule

SWE-Skills-Bench &
Source skills and benchmark reference material &
GitHub commit;
accessed 2026-05-06 &
Repository-level MIT License; upstream copyright notices and nested license
files preserved \\

Qwen model APIs &
Hosted LLM inference &
Provider API; model identifiers, decoding settings, and access dates recorded in
our run metadata &
Alibaba Cloud Model Studio / DashScope Terms of Service; no model weights
redistributed \\

DeepSeek model APIs &
Hosted LLM inference &
Provider API; model identifiers, decoding settings, and access dates recorded in
our run metadata &
DeepSeek Open Platform Terms of Service and DeepSeek Terms of Use; no model
weights redistributed \\

GLM model APIs &
Hosted LLM inference &
Provider API; model identifiers, decoding settings, and access dates recorded in
our run metadata &
Zhipu AI / BigModel Open Platform User Agreement and platform service terms; no
model weights redistributed \\

PyPI, npm, GitHub Actions, and Docker Hub metadata &
Public drift evidence, including versions, releases, workflow metadata,
repository metadata, and container metadata &
Accessed 2026-05-06; exact URLs, query timestamps, and retrieved metadata
snapshots recorded in our logs &
Public metadata accessed under the corresponding platform terms; no third-party
package source code, container images, or private contents redistributed \\

\dbname{} &
New benchmark, annotations, contract labels, evaluation metadata, documentation,
and release package &
Released by the authors; anonymized reviewer artifact for submission and
versioned public release upon acceptance &
CC BY 4.0 for benchmark data, labels, annotations, metadata, and documentation;
MIT License for evaluation code and scripts \\

\bottomrule
\end{tabular}
\end{table}

\paragraph{License files in the released artifact.}
The released artifact contains a top-level \texttt{LICENSE} file, a
\texttt{NOTICE} file, and an \texttt{ASSET\_LICENSES.md} inventory. The top-level
license states that, unless otherwise noted, code is licensed under the MIT
License and benchmark data, annotations, labels, metadata, and documentation are
licensed under CC BY 4.0. The asset inventory records each third-party source,
its pinned version or access date, its role in our experiments, and the
corresponding upstream license or provider terms.

\clearpage

\end{document}